\documentstyle[psfig,amssym]{l-aa}

\begin{document}
 
\thesaurus{12 (12.03.1; 12.03.3; 12.03.4; 12.12.1)}
 
\title{Gravitationally Lensed $\mu$Jy Radio Sources 
towards Galaxy Clusters}
\author{Asantha R. Cooray}
\institute{Department of Astronomy and Astrophysics, University of Chicago, 
Chicago IL 60637, USA. E-mail: asante@hyde.uchicago.edu}
\date{Received:  \today; accepted }
\maketitle


\begin{abstract} 
Galaxy clusters are expected to gravitationally lens 
background radio sources. However, due to
the smaller surface density of radio sources, when compared to
optical galaxies, such lensed 
events are rare. For an  example, 
 it is expected that there is no  lensed
radio source due to foreground galaxy clusters in the 1.4 GHz VLA FIRST survey.
However, at the $\mu$Jy level, the surface density of radio sources increases.
 Using the radio properties of the Hubble Deep Field (HDF) galaxies, 
we calculate the expected number of gravitationally lensed $\mu$Jy radio sources on the sky  due to foreground galaxy clusters for different 
cosmological models. For a flat cosmology with $\Omega_m=0.3$ and 
$\Omega_\Lambda =0.7$,
we predict $\sim$ 1500 lensed radio sources with flux densities $\sim$ 
10 to 1000
$\mu$Jy at 1.4 GHz. We discuss the possibility of detecting lensed 
$\mu$Jy radio
sources towards clusters with deep radio surveys.
Given the recent detection of a sub-mm selected lensed  $\mu$Jy radio 
source towards A370, it is suggested
that deep radio observations of  clusters should already contain such lensed
sources. Aided by amplification due to gravitational lensing, the search for 
lensed $\mu$Jy 
radio sources towards clusters are likely to recover  star-forming 
galaxies at redshifts of 1 to 3.

\end{abstract}


\section{Introduction}

It is now well known that gravitational lensing statistics is a useful
probe of the geometry of the universe, especially for the determination
of the cosmological constant.  In a recent paper (Cooray 1998; hereafter C98),
 we calculated the expected number
of gravitationally lensed arcs on the sky
 due to foreground galaxy clusters as a function of
cosmological parameters. The expected
number of arcs was calculated based on  the redshift distribution
of HDF galaxies (Williams et al. 1996), as determined by the photometric 
redshift
catalog of Sawicki et al. (1997), with an extrapolation to the whole sky. 
Similar to optical arcs, galaxy
clusters are expected to lens background radio sources. Such lensed
sources with high magnification should appear as arcs in
radio surveys. The number statistics
of lensed radio sources can be used to determine the cosmological parameters, to study the radio source evolution at high redshifts, and as discussed later,
properties of star forming galaxies at moderate to high redshifts.

The number statistics of lensed radio sources due to foreground clusters were first
calculated by Wu \& Hammer (1993). They predicted $\sim$ 10 lensed radio sources on the sky
down to a flux density limit of 0.1 mJy, and $\sim$ 100 lensed
radio sources down to 10 $\mu$Jy at 2.7 GHz (Figure~10 in Wu \& Hammer 1993).
At the source detection level of the VLA FIRST survey ($\sim$ 1 mJy; Becker, White \& Helfand 1995), there is only 
$\sim$ 2 to 3 lensed radio sources expected on the whole sky, 
and when compared to the area of the survey and its resolution, 
it is likely that there is no lensed source present. This prediction is compatible with
observational attempts to find lensed radio sources; Andernach, Gubanov \& Slee (1997)
searched the FIRST survey near Abell cluster cores and found no convincing candidates, 
and a statistical analysis of the radio positions towards clusters
showed no preferential tangential orientation, as expected from
gravitational lensing. Recently, a sub-mm selected source, SMM02399-0136,
towards cluster A370 was found to be lensed with an amplification of 2.5 (Ivison et al.
1997). The source was detected at 1.4 GHz, with a flux density of $\sim$ 525 $\mu$Jy.
This detection prompted us to calculate the expected number of lensed $\mu$Jy
sources present on the sky due to foreground clusters, and to refine the previous
predictions in Wu \& Hammer (1993). Since the predictions in Wu \& Hammer (1993) for sources 
down to mJy level are still expected to be valid, we will only concentrate on the $\mu$Jy sources
here. 

In \S~2 we describe our calculation and its inputs.  
In \S~3 we discuss the possibility of detecting lensed $\mu$Jy
sources. We follow the conventions that the Hubble
constant, $H_0$, is 100\,$h$\ km~s$^{-1}$~Mpc$^{-1}$, the present mean
density in the universe in units of the closure density is $\Omega_m$,
and the present normalized cosmological constant is $\Omega_\Lambda$.
In a flat universe, $\Omega_m+\Omega_\Lambda=1$.

\section{Expected Number of Lensed Sources}

In order to calculate the lensing rate for background $\mu$Jy sources
due to foreground galaxy clusters, we model the
lensing clusters as singular isothermal spheres (SIS) and use the
analytical  filled-beam approximation (see, e.g., Fukugita et al.
1992). Our calculation is similar to that of C98 in which we calculated the
expected number of lensed optical arcs on the sky due to foreground galaxy clusters (see, also, Cooray, Quashnock \& Miller 1998; hereafter CQM).
A main difference between the present paper and C98 is that we have not
 corrected for magnification bias in the present predictions (see, e.g., Kochanek 1991), 
primarily due to the lack of knowledge on the $\mu$Jy source luminosity function. Depending
on the shape of the luminosity function, or the slope of the number counts,
the predicted numbers will either increase or decrease due to magnification. The current consensus
on the slope of the $\mu$Jy counts (see, e.g., Windhorst et al. 1985) suggests
that there will be a slight excess in the number of sources towards clusters
due to gravitational lensing amplification.

In order to describe the background $\mu$Jy sources, we describe the
redshift and number distribution observed towards the HDF radio sources 
by Richards et al. (1998).
The main advantage in using the HDF data is the availability of redshift
information for $\mu$Jy sources. Also, HDF is one of the few areas where a deep
radio survey down to a flux limit of $\sim$ 2 $\mu$Jy at 1.4 GHz has been carried out.
The HDF contains 14 sources with flux densities of the order $\sim$ 6 to 500 $\mu$Jy at
8.5 GHz, and  11 of these sources have measured spectroscopic redshifts.
We converted the 8.5 GHz flux densities to 1.4 GHz using individual
spectral indices as presented by Richards et al. (1998). For  sources
with no measured spectral indices,  we assumed an index of
0.4, the mean spectral index observed for $\mu$Jy sources (Fomalont et al. 1991;
Windhorst et al. 1993; Richards et al. 1998). For the 3 sources with no 
measured spectroscopic redshifts, we used photometric redshifts from the
catalog of Fern\'andez-Soto, Lanzetta \&
Yahil (1998). 
We binned the redshift-number distribution in redshift steps of 0.25,
and calculated the lensing probability using filled-beam formalism. 
Similar to C98, we
calculated the $F$ parameter in lensing by describing the foreground lensing clusters using a Press-Schechter analysis.

We  calculated the expected number, $\bar N$,
of gravitationally lensed  radio sources on the sky as a function of
$\Omega_m$ and $\Omega_{\Lambda}$,
 and for a minimum amplification of $A_{\rm min}$ of 2
and 10 respectively. Since we are using the SIS model, the amplification is simply
equal to the ratio of length to width in observed lensing
arcs (see, e.g., Wu \& Mao 1996).
In Table 1, we list the expected number of strongly lensed arcs on the sky
for $A_{\rm min}=2$ and 10.

\begin{table}
\caption[]{Predicted number of lensed $\mu$Jy radio sources on the sky down to a flux density limit of
10 $\mu$Jy.}
\begin{flushleft}
\begin{tabular}{cccc}
\noalign{\smallskip}
\hline
\noalign{\smallskip}
$\Omega_m$ & $\Omega_\Lambda$ & $\bar N(A_{\rm min} \geq 2)$ &
$\bar N(A_{\rm min} \geq 10)$ \\
\noalign{\smallskip}
\hline
\noalign{\smallskip}
0.1 & 0.0 & 2155 & 25 \\
0.2 & 0.0 & 1303  & 16 \\
0.3 & 0.0 & 693 & 8 \\
0.4 & 0.0 & 376 & 4 \\
0.5 & 0.0 &  212  & 3 \\
0.6 & 0.0 & 123 & 1.5 \\
0.7 & 0.0 & 73 & 1 \\
0.8 & 0.0 & 46 & 0.5 \\
0.9 & 0.0 & 29 & 0.3  \\
1.0 & 0.0 & 19 & 0.2 \\
0.1 & 0.9 & 16465 & 163 \\
0.2 & 0.8 & 6697 &  65 \\
0.3 & 0.7 & 1581 & 16 \\
0.4 & 0.6 & 624 & 66 \\
0.5 & 0.5 & 326 & 36 \\
0.6 & 0.4 & 178 & 18 \\
0.7 & 0.3 & 103 & 11 \\
0.8 & 0.2 & 63 & 6 \\
0.9 & 0.1 & 40 & 3 \\
\noalign{\smallskip}
\hline
\end{tabular}
\end{flushleft}
\end{table}

\section{Discussion}

Using the redshift and flux distribution observed for HDF radio sources,
we have calculated the expected number of lensed $\mu$Jy sources
on the sky due to foreground clusters. By extrapolating the
observed properties towards the HDF to the whole sky, we have assumed 
that the HDF is a fair sample of the distant universe.
This assumption may be invalid given that the HDF was carefully selected
to avoid bright galaxies and radio sources. However, we have selected
to use the HDF data primarily because of the redshift
information for all $\mu$Jy sources detected, which is currently not available for other radio surveys with flux limits down to few $\mu$Jys. 

We have predicted $\sim$ 1500 lensed $\mu$Jy sources on the sky towards
clusters with X-ray luminosities greater than 
 $8 \times 10^{44} {\rm ergs\, s^{-1}}$,
for a cosmology with $\Omega_m=0.3$ and $\Omega_\Lambda=0.7$, consistent
with recent results based on lensing (CQM; Kochanek 1996), type Ia supernovae (Riess et al. 1998), and galaxy cluster baryonic fraction (Evrard 1997).  
The X-ray flux limit for foreground clusters
in our analysis is same as that of the clusters in the Le F\`evre et al. (1994) and Gioia \& Luppino (1994)
optical arc surveys, where 0.2 to 0.3 optical arc per cluster has been found down to a R band magnitude of $\sim$ 21.5.  
There are $\sim$ 7000 to 8000 such clusters on the whole sky (Bartelmann et al. 1997). We predict a similar, or slightly lower, rate for the $\mu$Jy sources, down to a flux density limit of 10 $\mu$Jy. 

We briefly describe the possibility of detecting such lensed sources
in deep radio surveys. 
Unlike optical surveys, radio surveys with interferometers such as VLA and
MERLIN are subjected to effects arising from instrumental limitations, primarily effects associated with resolution.  For example, there is a minimum and a maximum size for sources that can be detected and resolved with an interferometer. The largest angular scale to which the interferometer is sensitive restricts
the detection of high amplification sources, which are expected to appear as arcs, with  length to width ratios equal to amplification factors.
For the VLA A-array at 1.4 GHz, sources larger than $\sim$ 15$''$ may not likely
to be detected. Thus, observations of
radio arcs with length to width ratios greater than 10 may not easily be possible. In SIS model for gravitational lensing, most of the lensed
sources appear with amplification factors of 2 to 10. However, 
due to the convolution with synthesized beam,
ranging from $\sim$ 1$''$ to 5$''$, such sources are not likely to 
appear as arcs. Therefore,
detection of lensed sources with small amplifications 
are likely to be confused with foreground and cluster-member radio
sources, requiring a selection process to remove such confusing sources. 
Additional observations, especially optical identifications and redshifts 
may be required to establish the lensed nature of $\mu$Jy sources selected 
towards clusters. This is contrary to optical searches, where lensed
galaxies can easily be established due to the arc-like appearances.

In the present analysis, we have used the SIS models to calculate the lensing rate. However, certain galaxy clusters have complex potentials with bimodal
mass distributions. A careful analysis of the optical arcs and 
arclets towards A2218,
have shown that SIS models underestimate the lensing rate by as much as a factor of 2 (B\'ezecourt 1998). Thus, for certain clusters like A2218 and A370, it
is likely that there will be more lensed $\mu$Jy sources than the average prediction based on SIS models. By considering the ratio between observed number of optical arcs and arclets and the ratio of surface density of optical to $\mu$Jy sources, we find
that A2218 and A370, should contain 2 to 3 lensed $\mu$Jy sources down to 10 $\mu$Jy at 1.4 GHz. For A370, one such source has already been recovered (Ivison et al. 1997), through the sub-mm observations of Smail, Ivison \& Blain (1997). The VLA A-array 1.4 GHz data  (Owen \& Dwarakanath 1998, in preparation), 
in which the source was detected allows detection of sources down to a flux limit of 50 $\mu$Jy beam$^{-1}$ (5 $\sigma$). A quick analysis of
the same archival data suggests that there is at least one more
$\mu$Jy lensed source towards A370 (Cooray et al. 1998, in preparation). 
It is likely that  deep surveys of
galaxy clusters with MERLIN and VLA will allow detection of $\mu$Jy 
radio sources with amplifications of 2 to 10.

As discussed in Richards et al. (1998; see, also, Ivison et al. 1998), $\mu$Jy sources carry important information on the star formation rate and history.
Our lensing prediction 
suggests that most of the lensed sources are likely to occur between
redshifts of 1 and 3, with most around a redshift of 2.
Thus, observational searches for lensed sources will allow detection of 
moderate to high redshift
star-forming galaxies. The search for such galaxies
will be aided by the amplification due to gravitational lensing, allowing
detections of faint sources, below the flux limits of regular surveys.
It is likely that a careful analysis of lensed $\mu$Jy sources will allow 
the study of star formation at moderate to high redshift galaxies. 
As discussed in Richards et al. (1998), the low redshift $\mu$Jy sources, 
associated with spiral galaxies, are not likely to be found through clusters,
due to the low lensing rate. 
Based on our predictions and the detection of lensed
sources towards A370, we strongly recommend that deep observations of 
lensing clusters be carried out to find lensed sources and that such 
detections be followed up at other wavelengths.

\section{Summary}

Using the redshift and flux information for HDF radio sources and
a Press-Schechter analysis for clusters of galaxies,
we have calculated the expected number of lensed  $\mu$Jy 
sources towards galaxy clusters.
In a cosmology with $\Omega_m=0.3$ and $\Omega_\Lambda=0.7$,
we predict $\sim$ 1500 lensed sources towards clusters.
The radio emission associated with $\mu$Jy sources are expected
to be associated with star forming galaxies and search for such
lensed sources are likely to recover star forming galaxies at redshifts between 1 and 3. The possibility of detecting such lensed $\mu$Jy 
sources has already been
demonstrated by the recovery of a sub-mm selected galaxy at 1.4 GHz.
We suggest that similar deep VLA observations may already contain
lensed $\mu$Jy sources and that a careful analysis may be required to
establish the lensing nature of such sources.

\begin{acknowledgements}
I would like to acknowledge useful discussions and correspondences with Heinz
Andernach, Andr\'e Fletcher, Frazer Owen and Ian Smail on gravitational lensing of radio sources due to foreground clusters and the possibility of an observational search to find such sources.

\end{acknowledgements}

\end{document}